\documentclass[3p]{elsart}
\usepackage{graphicx,color}
\usepackage{amsmath}

\newcommand{\Slash}[1]{\ooalign{\hfil/\hfil\crcr$#1$}}

\begin{document}

\begin{frontmatter}
\title{Parity doubling of baryons in a chiral approach with three flavors}

\author[wroc]{Chihiro Sasaki}
\address[wroc]{Institute of Theoretical Physics,
University of Wroclaw, PL-50204, Wroclaw, Poland}

\begin{abstract}
We formulate a set of mass relations for the baryon
octet and decuplet with positive and negative parity in terms
of the order parameter of QCD chiral symmetry.
The Gell-Mann--Okubo mass formula and Gell-Mann's equal spacing
rule hold manifestly in this approach.
Thermal masses of the baryons are calculated in the mean
field approximation for various pion masses, and the results
are compared with the recent lattice studies.
A general trend of the nucleon, $\Delta$ and $\Omega$ parity-doublers
seen in the available lattice data can be understood qualitatively.
Expected mass modifications of other strange baryons are also given
with the physical and heavier pion masses.
\end{abstract}


\end{frontmatter}

\section{Introduction}
\label{sec:int}

Modifications of hadron properties in a hot/dense medium
have been explored as one of the key issues in the context
of QCD phase transition expected in heavy-ion collisions
and in the interior of compact stars~\cite{reviews}.
As chiral symmetry becomes restored, the hadron spectra with
opposite parity are expected to be degenerate. Yet, it remains
unclear to what extent they would influence over bulk thermodynamics
and experimental observables.

Recently, the first systematic study of thermal masses of
the octet and decuplet baryons with positive and negative parity
has been carried out in $N_f = 2+1$ flavored lattice QCD~\cite{Aarts}.
The temperature dependence of the nucleon, $\Delta$ and $\Omega$ masses
were extracted from temporal correlators, and they obviously exhibit
the parity doubling structure. The ground-state mass with positive parity
is rather stable against temperature, whereas the mass of the negative
parity partner drops substantially toward the chiral crossover
temperature. Although the simulations in \cite{Aarts}
have been performed with a relatively large pion mass,
$m_\pi \sim 400$ MeV, this is a clear signature of the partial
restoration of chiral symmetry in the baryonic sector.

In chiral approaches, a non-vanishing nucleon mass which stays finite
in chiral restored phase is introduced via so-called mirror assignment
of chirality in the parity doublet model~\cite{su2:mirror,Jido1,Jido2}.
The model has been applied to hot and dense baryonic matter and neutron
stars~\cite{Hatsuda,pdm,astro,astro2,SM,SLPR,PLRS1,BMS,su3:wrong,Weyrich,su3:n,su2:delta}
as well as the phenomenology in vacuum~\cite{su3:octet,Gallas,PLRS2,su3:vac}.

The two-flavored physics with parity doubling has been rather
extensively studied, whereas the studies with three flavors remain
quite limited. In particular, a systematic study of the in-medium
masses of the octet and decuplet states is still missing.
In this paper, we start with the general SU(3) Lagrangian and
deduce a complete set of the mass relations in the parity doubling scenario,
in a manifestly consistent manner with the celebrated Gell-Mann--Okubo
mass formula and Gell-Mann's equal spacing rule.
We also study the thermal behavior of the baryon masses in
a self-consistent chiral approach under the mean field approximation.
For qualitative comparison to the lattice data~\cite{Aarts},
we demonstrate the calculations with the physical and heavier pion masses.

\section{Octet and decuplet baryons}
\label{sec:eff}

Introducing an octet $g_8$ and a singlet $g_1$ coupling constants,
the general SU(3) interaction Lagrangian with a meson field $\Phi$
is given by~\cite{deSwart,Stoks}
\begin{eqnarray}
&&
{\mathcal L}_{BB\Phi}
=
-\sqrt{2}g_8\alpha\mbox{tr}\left[\bar{B}\left[\Phi, B\right]\right]
\nonumber\\
&&
{}- \sqrt{2}g_8 (1-\alpha)\left(
\mbox{tr}\left[\bar{B}\left\{\Phi, B\right\}\right]
{}- \frac{2}{3}\mbox{tr}\left[\bar{B}B\right]\mbox{tr}\left[\Phi\right]
\right)
\nonumber\\
&&
{}- \frac{g_1}{\sqrt{3}}\mbox{tr}\left[\bar{B}B\right]
\mbox{tr}\left[\Phi\right]\,,
\label{lag:stoks}
\end{eqnarray}
where $\alpha$ is known as the $F/(F+D)$ ratio.
Masses of the baryon octet are generated when an octet of scalar fields
get condensed~\footnote{
 For details of the scalar vacuum expectation matrix staying invariant
 via a nonlinear transformation, see Appendix of \cite{Stoks}.
}, and {\it all of them} depend on the light-quark $\sigma_q$ and the strange-quark
condensates $\sigma_s$. As suggested in \cite{Papa2}, there exists a special
set of the parameters, $(\alpha, g_1) = (1, \sqrt{6}g_8)$, which leads to
the nucleon mass depending only on the $\sigma_q$.

The extension to the parity doublet picture is carried out
following \cite{su3:octet}. The chiral invariant mass $m_{0}$ of
two fermions $\psi_1$ and $\psi_2$ is introduced as
\begin{equation}
{\mathcal L}_{\rm inv}
= m_{0}\mbox{tr}\left[\bar{\psi}_1\gamma_5\psi_2
{}- \bar{\psi}_2\gamma_5\psi_1\right]\,.
\end{equation}
In the physical basis, the nucleon masses with positive and
negative parity are found as
\begin{equation}
m_{N_\pm}
= \sqrt{\alpha_N^2\sigma_q^2 + m_{0}^2} \mp \beta_N\sigma_q\,,
\label{pd1}
\end{equation}
where $\alpha_N$ and $\beta_N$ are the nucleon coupling constants to
the scalar mesons.
The decuplet baryons are introduced as the Rarita-Schwinger fields,
and the delta masses are given in a similar fashion~\cite{Jido1,Jido2}:
\begin{equation}
m_{\Delta_\pm}
= \sqrt{\alpha_\Delta^2\sigma_q^2 + m_{0}^2} \mp \beta_\Delta\sigma_q\,,
\label{pd2}
\end{equation}
with constants $\alpha_\Delta, \beta_\Delta$ and $m_{0}$~\footnote{
 The invariant mass for the octet state can be different from
 that for the decuplet. In this work, we take a common value for simplicity.
 In fact, the two invariant masses are found to be rather close
 in the recent lattice study~\cite{Aarts}.
 See also the discussion in the Conclusions section.
}.

In extending the above to the other octet and decuplet baryons
with a chiral invariant mass, we encounter a problem.
Because of the nonlinear $\sigma$ dependence in Eq.~(\ref{pd1}),
the Gell-Mann--Okubo mass relation for the baryon octet,
\begin{equation}
\frac{3}{4}m_\Lambda + \frac{1}{4}m_\Sigma
{}- \frac{1}{2}\left(m_N + m_\Xi\right) = 0\,,
\label{GMO}
\end{equation}
is violated, unless the exact SU(3) limit is taken.
Another non-trivial relation for the decuplet
baryon, Gell-Mann's equal spacing rule,
\begin{equation}
m_{\Sigma^\ast} - m_\Delta = m_{\Xi^\ast} - m_{\Sigma^\ast}
= m_\Omega - m_{\Xi^\ast}\,,
\label{GM}
\end{equation}
is not satisfied either~\footnote{
 Therefore, the parameterization for the baryon parity doublers
 given in \cite{su3:wrong,astro2} does not reproduce
 the low-energy relations~(\ref{GMO}) and (\ref{GM}).
}.
We note that, without $m_0$, the two relations, (\ref{GMO}) and (\ref{GM}),
hold for any $\alpha, g_1$ and $g_8$ when the Lagrangian ~(\ref{lag:stoks})
is used.

Therefore, we shall adopt the following mass relations for
the octet parity doublers
\begin{eqnarray}
m_{N_\pm}
&=&
\left(a_N \mp b_N\right)3\sigma_q + m_{0}\,,
\nonumber\\
m_{\Sigma_\pm}
&=&
\left(a_N \mp b_N\right)\left(2\sigma_q + \sqrt{2}\sigma_s\right)
{}+ m_{0} + m_1\,,
\nonumber\\
m_{\Lambda_\pm}
&=&
\left(a_N \mp b_N\right)\left(2\sigma_q + \sqrt{2}\sigma_s\right)
{}+ m_{0} + m_3\,,
\nonumber\\
m_{\Xi_\pm}
&=&
\left(a_N \mp b_N\right)\left(\sigma_q + 2\sqrt{2}\sigma_s\right)
{}+ m_{0} + m_2\,,
\label{mass:octet}
\end{eqnarray}
where three parameters $m_{1,2,3}$ are introduced in order to
generate a mass difference between the $\Sigma$ and $\Lambda$ due to
the explicit chiral symmetry breaking.
They are related via the Gell-Mann--Okubo relation~(\ref{GMO}) as
\begin{equation}
m_1 = 2m_2 - 3m_3\,.
\label{m1}
\end{equation}
The decuplet parity doublers follow
\begin{eqnarray}
m_{\Delta_\pm}
&=&
\left(a_\Delta \mp b_\Delta\right)3\sigma_q + m_{0}\,,
\nonumber\\
m_{\Sigma^\ast_\pm}
&=&
\left(a_\Delta \mp b_\Delta\right)\left(2\sigma_q + \sqrt{2}\sigma_s\right)
{}+ m_{0} + m_s\,,
\nonumber\\
m_{\Xi^\ast_\pm}
&=&
\left(a_\Delta \mp b_\Delta\right)\left(\sigma_q + 2\sqrt{2}\sigma_s\right)
{}+ m_{0} + 2m_s\,,
\nonumber\\
m_{\Omega_\pm}
&=&
\left(a_\Delta \mp b_\Delta\right)3\sqrt{2}\sigma_s
{}+ m_{0} + 3m_s\,,
\label{mass:decuplet}
\end{eqnarray}
where the terms including $m_s$ are added for relatively strong explicit
symmetry-breaking in the strange-quark sector in such a way that all is
consistent with the equal spacing rule~(\ref{GM}). We will use
the current strange-quark mass $m_s = 0.1$ GeV~\cite{PDG}.
One readily sees that the low-energy relations (\ref{GMO}) and (\ref{GM})
are now satisfied even in the presence of explicit SU(3) breaking.

The above mass relations~(\ref{mass:octet}) and (\ref{mass:decuplet})
can be deduced from Eqs.~(\ref{pd1}) and (\ref{pd2})
by assuming $\sigma_{q,s} \ll m_{0}$. As discussed in the
Introduction, our main attention will be put to the mass modifications
near chiral symmetry restoration, where in-medium condensates are
certainly smaller than their vacuum values.
In fact, its order of magnitude extracted from the lattice results~\cite{Aarts}
is compatible to the vacuum nucleon mass with positive parity.
Thus, the limiting case
$\sigma_{q,s} \ll m_{0}$ is well justified even at very low temperature, and the mass
relations are now fully consistent with the model-independent
low-energy theorems (\ref{GMO}) and (\ref{GM})~\footnote{
 At $T \sim 0$, the negative-parity nucleon can be integrated out, so that the resultant
 Lagrangian includes only the positive-parity nucleon and mesons. Non-linear realization
 of chiral symmetry allows the fermion mass-operator $\bar{\psi}\psi$. When one introduces
 a term $m_0\bar{N}N$ on top of the meson-nucleon interaction in the Lagrangian, one finds
 the entire nucleon mass in the form of $m_N = c_1\sigma_q + m_0$ with a certain constant $c_1$.
}.
We note that the approximated expression (\ref{mass:octet}) leads to
a quite similar behavior in the thermal nucleon masses to that obtained from
the original non-linear form with respect to the condensate,
Eq.~(\ref{pd1}).
In particular, temperature dependence of the mass difference of the nucleon parity-doublers is identical.
Thus, the two expressions describe the physics
equally well in the range of temperature of interest. This encourages us to apply the same
scheme to the strange baryons.

We emphasize that the value taken from the lattice study~\cite{Aarts} is not the conclusive number.
However, such a large $m_0$ is actually consistent with the earlier model studies
applied to the vacuum and to nuclear matter~\cite{Gallas,pdm}, where the $m_0$ values were
determined to optimize the known phenomenological properties of the systems.
The smallest value found in the available papers is of order $\Lambda_{\rm QCD}$~\cite{su2:mirror}.
Thus, the expansion in the ratio
$\sigma_q/m_0 \sim 92\,\mbox{MeV}/\Lambda_{\rm QCD} < 1$ is still adequate.
Although the thermal profiles of the baryon masses will certainly change
according to how large the $m_0$ is, the normalized
mass difference is unaffected, as emphasized.

The parameters in Eqs.~(\ref{mass:octet}) and (\ref{mass:decuplet})
at zero temperature are determined as in Table~\ref{paraB}
where the following input was used~\footnote{
 A different assignment is possible; one can chose e.g. $N(1650)$ as the negative-parity partner
 of the lowest nucleon. Such a variation in the assignment does not yield any significant difference
 in the bulk equation of state, fluctuations and correlations~\cite{Morita}.
};
$m_{N_+} = 0.939$ GeV, $m_{N_-} = 1.535$ GeV,
$m_{\Sigma_+} = 1.193$ GeV, $m_{\Xi_+} = 1.318$ GeV,
$m_{\Delta_+} = 1.232$ GeV, $m_{\Delta_-} = 1.710$ GeV,
$m_{\Sigma^\ast_+} = 1.383$ GeV, and
the pion and kaon decay constants $f_\pi = \sigma_q = 92.4$ MeV,
$f_K = (f_\pi + \sqrt{2}\sigma_s)/2 = 113$ MeV~\cite{PDG}.
\begin{table}
\begin{center}
\begin{tabular*}{12cm}{@{\extracolsep{\fill}}ccccccc}
\hline
$a_N$ & $b_N$ & $m_1$ [GeV] & $m_2$ [GeV] &
$a_\Delta$ & $b_\Delta$ & $m_{0}$ [GeV]
\\
\hline
$1.26$ & $1.08$ & $0.247$ & $0.364$ &
$2.10$ & $0.862$ & $0.889$
\\
\hline
\end{tabular*}
\end{center}
\caption{
Set of parameters in the baryon-mass relations.
}
\label{paraB}
\end{table}
This leads to the masses of the remaning octet states,
\begin{eqnarray}
&&
m_{\Lambda_+} = 1.11\,\mbox{GeV}\,,
\quad
m_{\Lambda_-} = 1.79\,\mbox{GeV}\,,
\nonumber\\
&&
m_{\Sigma_-} = 1.88\,\mbox{GeV}\,,
\quad
m_{\Xi_-} = 2.09\,\mbox{GeV}\,,
\end{eqnarray}
and of the other decuplet states,
\begin{eqnarray}
&&
m_{\Sigma^\ast_-} = 1.93\,\mbox{GeV}\,,
\nonumber\\
&&
m_{\Xi^\ast_+} = 1.53\,\mbox{GeV}\,,
\quad
m_{\Xi^\ast_-} = 2.15\,\mbox{GeV}\,,
\nonumber\\
&&
m_{\Omega_+} = 1.69\,\mbox{GeV}\,,
\quad
m_{\Omega_-} = 2.38\,\mbox{GeV}\,.
\end{eqnarray}
These masses of the positive-parity states are
in quite good agreement with the PDG values.
Masses, spin and parity of the above negative-parity states
are not fully confirmed in experiments, thus they are excluded in
the PDG Summary Table.

\section{Effective masses in hot matter}
\label{sec:mf}

The mass modifications will be brought by the quark condensates
$\sigma_q$ and $\sigma_s$ in a medium.
To quantify those effects, we take the standard linear sigma
model Lagrangian with three flavors:~\footnote{
 Since Eqs.~(\ref{mass:octet}) and (\ref{mass:decuplet}) are model-independent
 at tree level, they are also true in a quark-meson model as considered in this section.
 The in-medium condensates can be computed in any alternative approaches.
}
\begin{eqnarray}
{\mathcal L}_{\rm L}
&=&
\bar{q}\left(i\Slash{\partial} - g T^a\left(
\sigma^a + i\gamma_5\pi^a
\right)\right)q
\nonumber\\
&&
{}+ {\rm tr}\left[\partial_\mu\Sigma^\dagger\cdot\partial^\mu\Sigma\right]
{}- V_{\rm L}(\Sigma)\,,
\end{eqnarray}
where the potential, including $U(1)_A$ breaking effects, is
\begin{eqnarray}
V_{\rm L}
&=&
m^2{\rm tr}\left[\Sigma^\dagger\Sigma\right]
{}+ \lambda_1\left({\rm tr}\left[\Sigma^\dagger\Sigma\right]\right)^2
\nonumber\\
&&
{}+\lambda_2{\rm tr}\left[\left(\Sigma^\dagger\Sigma\right)^2\right]
{}- c\left(\det\Sigma + \det\Sigma^\dagger\right)
\nonumber\\
&&
{}- {\rm tr}\left[h\left(\Sigma + \Sigma^\dagger\right)\right]\,,
\end{eqnarray}
with the chiral field $\Sigma = T^a\Sigma^a
= T^a\left(\sigma^a + i\pi^a\right)$ as a $3 \times 3$ complex matrix
in terms of the scalar $\sigma^a$ and the pseudoscalar $\pi^a$ states.
The last term with $h = T^a h^a$ breaks the chiral symmetry explicitly.

For thermodynamic calculations, we employ the mean field approximation.
We also assume that there is the SU(2) isospin symmetry in the up and
down quark sector.
This leads to $\sigma_0$ and $\sigma_8$ as non-vanishing condensates,
which contain both strange and  non-strange components.
The pure non-strange and strange parts are obtained through the following
rearrangement,
\begin{equation}
\begin{pmatrix}
\sigma_q\\
\sigma_s
\end{pmatrix}
=
\frac{1}{\sqrt{3}}
\begin{pmatrix}
\sqrt{2} & 1 \\
1 & -\sqrt{2}
\end{pmatrix}
\begin{pmatrix}
\sigma_0 \\
\sigma_8
\end{pmatrix}\,.
\end{equation}
In this basis, the effective quark masses read
\begin{equation}
M_q
=
\frac{g}{2}\sigma_q\,,
\quad
M_s = \frac{g}{\sqrt{2}}\sigma_s\,.
\end{equation}
The explicit symmetry breaking terms are related with
the pion and kaon masses as
\begin{equation}
h_q = f_\pi m_\pi^2\,,
\quad
h_s = \sqrt{2}f_K m_K^2 - \frac{f_\pi m_\pi^2}{\sqrt{2}}\,.
\label{exp}
\end{equation}

The entire thermodynamic potential is given by
\begin{eqnarray}
\Omega
= \Omega_q + V_{\rm L}\,,
\end{eqnarray}
with the thermal-quark contribution
\begin{eqnarray}
\Omega_q
&=&
6 T\sum_{f=u,d,s}
\int\frac{d^3p}{(2\pi)^3}
\left[\ln\left(1-n_f\right)
{}+ \ln\left(1-\bar{n}_f\right)\right]\,,
\nonumber\\
\end{eqnarray}
with the Fermi-Dirac distribution functions,
$n_f, \bar{n}_f = 1/\left(1 + e^{(E_f \mp \mu_f)/T}\right)$,
and the quasi-quark energies,  $E_f = \sqrt{p^2 + M_f^2}$.
By minimizing the thermodynamic potential, the two mean-fields
are determined self-consistently at a given $T$ and $\mu_f$ via
$
\frac{\partial\Omega}{\partial\sigma_q}
= \frac{\partial\Omega}{\partial\sigma_s}
= 0
$\,.
In this work, we consider thermodynamics at $\mu_f=0$,
and use the model parameters fixed in the vacuum~\cite{su3:LSM},
summarized in Table~\ref{paraL}.
\begin{table}
\begin{center}
\begin{tabular*}{12cm}{@{\extracolsep{\fill}}ccccccc}
\hline
$c$ [GeV] & $m$ [GeV] & $\lambda_1$ & $\lambda_2$ & $h_q$ [GeV$^3$]
& $h_s$ [GeV$^3$] & $g$ \\
\hline
$4.81$ & $0.343$ & $1.40$ & $46.48$ & $(0.121)^3$ & $(0.336)^3$
& $6.5$ \\
\hline
\end{tabular*}
\end{center}
\caption{
Set of parameters in the light sector with
$m_\sigma = 600$ MeV~\cite{su3:LSM}.
}
\label{paraL}
\end{table}

\subsection{Condensates}

\begin{figure}
\begin{center}
\includegraphics[width=6.5cm]{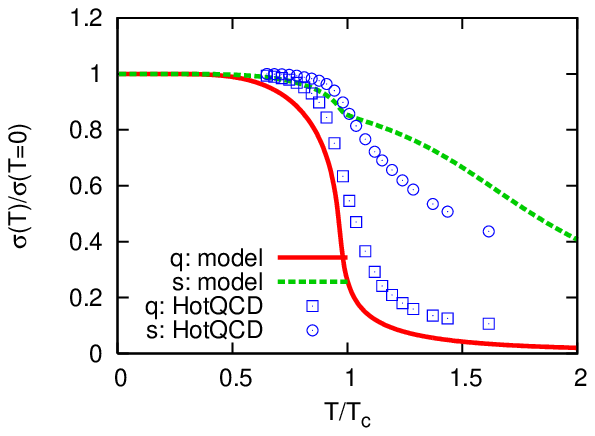}
\includegraphics[width=6.5cm]{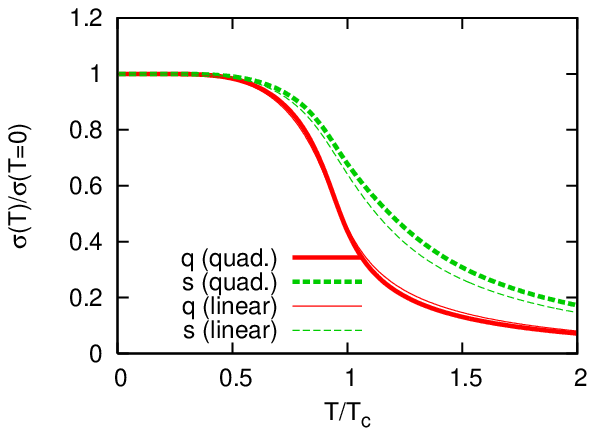}
\caption{
Thermal expectation values of the mean fields,
$\sigma_q$ and $\sigma_s$, calculated in the chiral model
for the physical $m_\pi$ and $m_K$ (left) and for heavier
meson masses with a quadratic $r = (m_\pi/m_K)^2 = 0.64$
and a linear $r = m_\pi/m_K = 0.8$ ratios (right).
The pseudo-critical temperatures fixed
from the chiral susceptibility are
$T_c = 151$ MeV for the physical $r$,
$T_c = 203$ MeV for $r = 0.64$ and
$T_c = 272$ MeV for $r = 0.8$, respectively.
On the left-hand figure, the corresponding lattice data
for the physical pion and kaon masses~\cite{HotQCD}
are given for comparison.
}
\label{fig:sigma}
\end{center}
\end{figure}

In-medium condensates $\sigma_{q}$ and $\sigma_s$ at finite
temperature are shown in Fig.~\ref{fig:sigma} (left).
It is clearly seen that the melting strange condensate
is delayed because of the stronger explicit symmetry
breaking with the strange quark. Nevertheless, the
$\sigma_s$ exhibits an abrupt, milder than the $\sigma_q$ though,
change near the crossover temperature $T_c$, which is driven
by the light flavor chiral dynamics. For comparison, we also
show the corresponding lattice date taken from \cite{HotQCD}
where the thermodynamic quantities were calculated in
the physical pion and kaon masses. The light-quark condensate
follows more or less the model result, whereas the strange-quark
condensate shows a rather mild behavior but the trend seen in
the model calculation stays.
We will not make any extrapolation to higher temperature using the model
since the validity of this sort of hadronic models is questionable.

In the recent lattice study~\cite{Aarts}, their simulations were
carried out with a relatively large up- and down-quark masses
and the physical strange quark mass, leading to $m_\pi \sim 400$ MeV.
Thus, it is constructive to study the condensates also for a heavier
quark mass. To this end, we introduce the mass ratio,
\begin{equation}
r = \left(\frac{m_\pi}{m_K}\right)^2\,,
\end{equation}
which scales like the quark mass ratio $m_q/m_s$ guided by
the Gell-Mann--Oakes--Renner (GOR) relation. The physical value is
$r_{\rm phys} = 0.077$ with $m_\pi = 138$ MeV and
$m_K = 496$ MeV~\cite{PDG}. Since this is considerably smaller
than the above-mentioned lattice setup $r_{\rm lat} \sim 0.64$,
the quadratic scaling imposed by the GOR relation might be violated
in the system with a very massive $m_\pi$.
We therefore examine the quark-mass dependence assuming a linear
scaling, $r = m_\pi/m_K$, as well, by replacing $m_\pi^2$ with $r m_K$
in Eq.~(\ref{exp}).
The results are summarized in
Fig.~\ref{fig:sigma} (right). It is found that a somewhat stronger
deviation from the quadratic scaling is seen in the $\sigma_s$
than in the $\sigma_q$, but the difference is rather minor.
Thus, in the following, we will consider only the quadratic case.

Note that the Gell-Mann--Okubo mass formula~(\ref{GMO}) and
Gell-Mann's equal spacing rule (\ref{GM}) hold
at any temperature.

\subsection{Baryon octet}

By substituting the obtained in-medium condensates into the mass
relations (\ref{mass:octet}) and (\ref{mass:decuplet}),
all the baryon masses are now obtained.
We emphasize that chiral symmetry restoration does not dictate
directly how the masses of the nucleon and baryon resonances
should go. What is required for the mass spectra is that the parity
partners become degenerate. Namely, the mass difference between the
positive and negative parity states should vanish when the chiral
symmetry is fully restored, and this is a secure model-independent
statement. We therefore introduce the following quantity for the
parity doublers:
\begin{equation}
\frac{\delta m(T)}{\delta m(T=0)} = \frac{m_-(T) - m_+(T)}{m_-(T=0) - m_+(T=0)}\,.
\end{equation}

\begin{figure}
\begin{center}
\includegraphics[width=6.5cm]{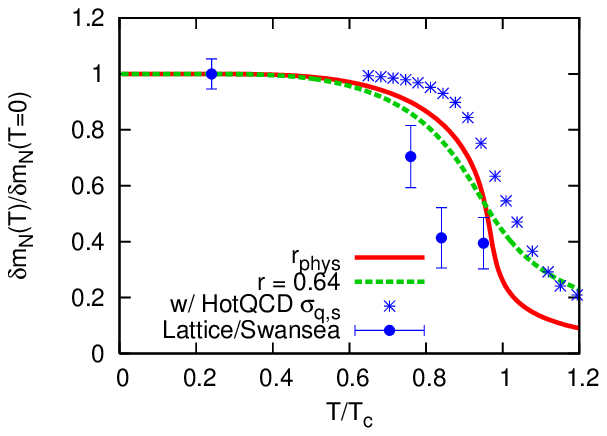}
\includegraphics[width=6.5cm]{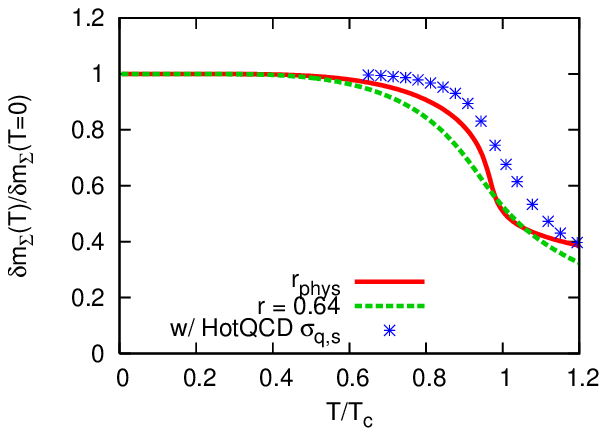}
\includegraphics[width=6.5cm]{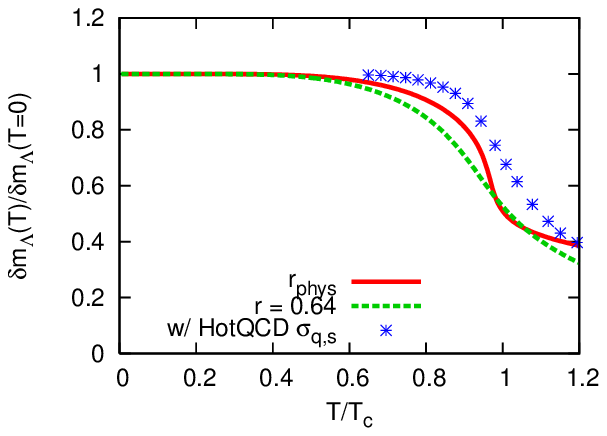}
\includegraphics[width=6.5cm]{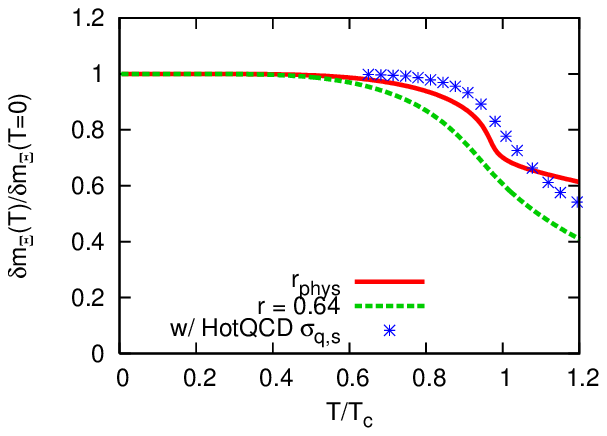}
\caption{
Temperature dependence of the ratio $\delta m$ for
the baryon octet with the physical $r$ and $r = 0.64$.
The stars are calculated by using the thermal profiles of
the $\sigma_q$ and $\sigma_s$ taken from lattice
simulations with the physical setup~\cite{HotQCD}.
The filled circles of $\delta m_N$ are the lattice data
with $r \sim 0.64$~\cite{Aarts}.
}
\label{fig:octet}
\end{center}
\end{figure}

The thermal mass difference of the baryon octet is presented in
Fig.~\ref{fig:octet}. For comparison, we also show the masses
calculated with the thermal profiles $\sigma_{q,s}$ shown in
Fig.~\ref{fig:sigma} (left).
The $\delta m_N$ evolves with the thermal $\sigma_q$ and
drops substantially toward $T_c$. It agrees well with the
result calculated with the lattice $\sigma_{q}$.
The trend becomes milder due to stronger explicit symmetry breaking
when the pion mass is increased. One sees a fairly good
agreement with the $\delta m_N$ from lattice QCD with heavier
pion~\cite{Aarts}.

Different behavior of the $\sigma_s$ from the $\sigma_q$ comes in to
the states including strangeness, $\Sigma, \Lambda$ and $\Xi$.
There is a clear contrast to the $\delta m_N$:
The mass difference $\delta m$ of the strange baryons is {\it reduced}
when the ratio $r$ is increased, despite a stronger explicit
symmetry breaking. This is because the underlying flavor symmetry
turns into an SU(3) when the ratio $r$ approaches unity,
which is seen already in Fig.~\ref{fig:sigma}.

\subsection{Baryon decuplet}

\begin{figure}
\begin{center}
\includegraphics[width=6.5cm]{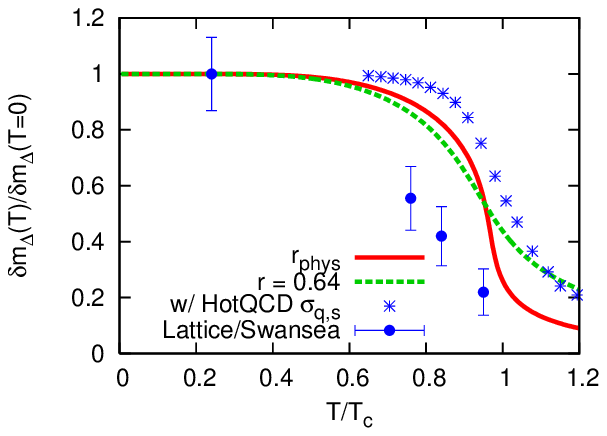}
\includegraphics[width=6.5cm]{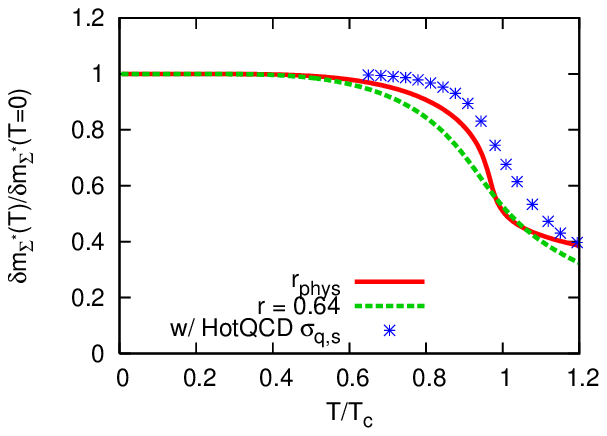}
\includegraphics[width=6.5cm]{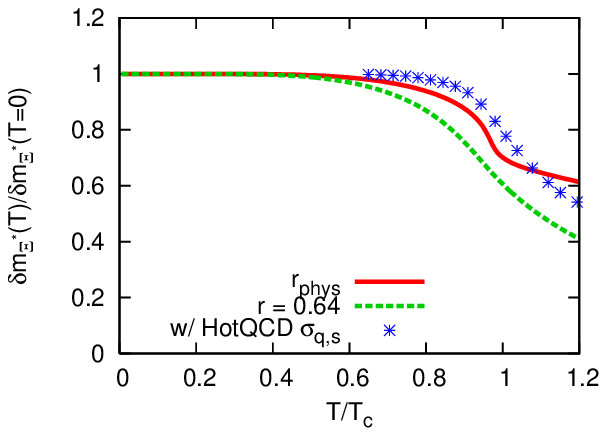}
\includegraphics[width=6.5cm]{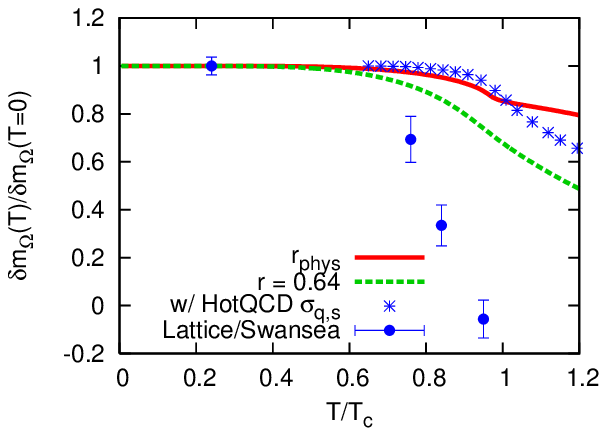}
\caption{
The same as in Fig.~\ref{fig:octet}, but for
the baryon decuplet.
}
\label{fig:decuplet}
\end{center}
\end{figure}

The mass modifications of the baryon decuplet in Fig.~\ref{fig:decuplet}
have a quite similar trend to those of the baryon octet.

We see a rather strong deviation in the $\delta m_\Omega$
with $r = 0.64$ from the corresponding lattice points.
An approximate SU(3) structure for larger $r \sim 1$
supports this tendency to some extent. Yet, there remains
a qualitative difference. The actual mass may not follow
the linear dependence as in Eq.~(\ref{mass:decuplet}).
It requires more realistic treatment
of the in-medium strange baryons to resolve this discrepancy.
One missing piece in the current hadronic approach is
a mechanism of deconfinement. It is interesting to see
in a more microscopic model how much the onset of deconfinement
affects the hadronic quantities slightly below $T_c$.

\section{Conclusions}
\label{sec:conc}

We have formulated a complete set of mass relations for the baryon
octet and decuplet with positive and negative parity in a chiral
approach. The celebrated Gell-Mann--Okubo mass formula and
Gell-Mann's equal spacing rule are now manifest.
We have also demonstrated the thermodynamic calculations with several
pion masses in the mean field approximation, and have shown thermal
baryon masses in terms of the approximate order parameter of QCD
chiral symmetry.

For the physical pion and kaon masses, the mass splitting $\delta m$
between the positive and negative parity states crucially depends on
their strange-quark content, which is clearly seen in the numbers
the $\sigma_s$ in the mass relations. The $\delta m_N$ and
$\delta m_\Delta$ exhibit an abrupt drop near the chiral crossover
temperature $T_c$, whereas the $\delta m$ of strange baryons
drops rather slowly. The size of the $\delta m$ grows gradually
with strangeness.

For a qualitative comparison to the recent lattice results~\cite{Aarts},
we have studied the thermodynamics with a larger $m_\pi$-to-$m_K$ ratio
$r$ than its physical value. Given $r$ comparable to the lattice setup
of \cite{Aarts}, it is clearly seen that the $\sigma_s$ does not differ
much from the $\sigma_s$ because of the pion mass rather closer to the
kaon mass. Thus, the underlying flavor symmetry is an approximate SU(3).
Consequently, the $\delta m$s of the strange baryons are more reduced
for $r$ closer to unity (the exact SU(3) limit). The $\delta m$s of
$\Sigma, \Lambda, \Xi, \Sigma^\ast$ and $\Xi^\ast$ are to be compared
with future lattice results when available.

One of the immediate questions is how we see such thermal modifications
of the baryons masses in bulk thermodynamic quantities and observables.
These resonances get broadened because of the medium effects, and in fact
the importance of the resonance widths have been studied in the context of
the fluctuations of conserved changes and the pion distributions in
heavy-ion collisions~\cite{kielce,kappa,rhowidth}.
Since the width broadening and the mass modification should be linked,
a more realistic next-step would be in line with the Greens function
method with a proper extension of the work done in \cite{WFN}.

The lattice results~\cite{Aarts} also show that the chiral-invariant
mass for the nucleon is quite close to that of the $\Delta$ state.
In general, they can differ due to e.g. the spin-spin interaction.
These survival masses should be saturated by the condensate of
the chiral-even operators, in particular, the gluon condensate is
a promising major contributor~\cite{SLPR,BMS}.
In hadronic approach, the gluon condensate is introduced as a dilaton
associated with the conformal symmetry breaking.
When it is applied to a hot/dense medium,
the vacuum expectation value (VEV) of the dilaton gives the non-vanishing
value of $m_0$ at a given temperature or density. Under the mean field approximation, however,
the VEV is found not to change much up to the chiral crossover, reflecting the fact that
the sigma boson is much lighter than the glueball~\cite{BMS}. Therefore, as the first approximation,
a frozen dilaton, leading to a constant $m_0$, is fairly acceptable. A nearly-constant $m_0$
may imply that it is dominated by the color-magnetic gluons, rather than the color-electric
component which drives deconfinement.
Since this issue is
ultimately linked to the fundamental question, what the origin of
mass is, further studies at finite temperature and density will shed more light on the dynamical
generation of the hadron masses.

\subsection*{Acknowledgments}

We acknowledge stimulating discussions with K.~Redlich.
This work has been partly supported
by the Polish Science Foundation (NCN) under
Maestro grant 2013/10/A/ST2/00106.


\end{document}